\documentclass[aps,prl,twocolumn,groupedaddress,showpacs,preprintnumbers,
amsmath,amssymb]{revtex4}
\usepackage[dvips]{graphicx}
\usepackage{amsfonts}
\usepackage{amssymb}
\usepackage{float}
\usepackage{natbib}
\usepackage{amsmath}
\usepackage{float}
\usepackage{dcolumn}
\usepackage{lscape}

\newcommand{\eb}{\bar{E}}
\newcommand{\et}{\widetilde{E}} 
\newcommand{\etsb}{\widetilde{E}_{\scriptscriptstyle SB}}

\newcommand{\rhob}{\bar{\rho}}
\newcommand{\rhot}{\widetilde{\rho}}
\newcommand{\rct}{{\cal R}_{\scriptscriptstyle T}}

\newcommand{\nc}{{\cal N}}
\newcommand{\ncb}{\bar{\cal N}} 
\newcommand{\nct}{\widetilde{\cal N}}
\newcommand{\mub}{\bar{\mu}} 

\newcommand{\rhotI}{\widetilde{\rho}_{\scriptscriptstyle I}}
\newcommand{\rhotT}{\widetilde{\rho}_{\scriptscriptstyle T}}

\newcommand{\etS}{\widetilde{E}_{\scriptscriptstyle I}} 
\newcommand{\etA}{\widetilde{E}_{\scriptscriptstyle T}}

\newcommand{\etmean} {\langle \widetilde{E} (N) \rangle_{\scriptscriptstyle N}} 
\newcommand{\etsbmean}{\langle \widetilde{E}_{\scriptscriptstyle SB}
(N) \rangle_{\scriptscriptstyle N}}

\makeatletter
\newlength{\earraycolsep}
\def\eqnarray{\stepcounter{equation}\let\@currentlabel%
\theequation
\global\@eqnswtrue\m@th
\global\@eqcnt\z@\tabskip\@centering\let\\\@eqncr
$$\halign to\displaywidth\bgroup\@eqnsel\hskip\@centering
$\displaystyle\tabskip\z@{##}$&\global\@eqcnt\@ne
\hskip 2\earraycolsep \hfil$\displaystyle{##}$\hfil
&\global\@eqcnt\tw@ \hskip 2\earraycolsep
$\displaystyle\tabskip\z@{##}$\hfil
\tabskip\@centering&\llap{##}\tabskip\z@\cr}
\makeatother

\newenvironment{narrow}[2]{%
\begin{list}{}{%
\setlength{\topsep}{0pt}%
\setlength{\leftmargin}{#1}%
\setlength{\rightmargin}{#2}%
\setlength{\listparindent}{\parindent}%
\setlength{\itemindent}{\parindent}%
\setlength{\parsep}{\parskip}}%
\item[]}{\end{list}}

\begin{document}

\title{On the ground--state energy of finite Fermi systems}

\author{J\'er\^ome Roccia and Patricio Leboeuf}

\affiliation{
Laboratoire de Physique Th\'eorique et Mod\`eles Statistiques,
CNRS, B\^at. 100, \\ Universit\'e de Paris-Sud, 91405 Orsay Cedex, France\\}

\date{\today}

\begin{abstract} 
  We study the ground--state shell correction energy of a fermionic gas in a mean--field approximation.
  Considering the particular case of 3D harmonic trapping potentials,
  we show the rich variety of different behaviors (erratic, regular, supershells) that appear 
  when the number--theoretic properties of the frequency ratios are varied. For self--bound systems,
  where the shape of the trapping potential is determined by energy minimization,
  we obtain accurate analytic formulas for the deformation and the shell correction energy 
  as a function of the particle number $N$.
  Special attention is devoted to the average of the shell correction energy. We explain
  why in self--bound systems it is a decreasing (and negative) function of $N$.
\end{abstract}



\maketitle

\section{I. Introduction}

When estimating the ground--state energy of interacting Fermi gases, one can
schematically rely on two different approaches. For few--particle systems (say,
less than 10-20), direct {\sl ab initio} calculations are
available at present \cite{wiringa}. For larger systems, and in particular for global
mass calculations in atomic nuclei, approximate schemes based either on
mean--field calculations \cite{rs} or shell models \cite{rs, bm} are inevitable.
In mean field theories, in which we focus here, the energy or mass of the
system is naturally decomposed into a smooth part and a fluctuating part.
Methods like Thomas-Fermi (TF) or Wigner-Kirkwood theories provide an
approximation to the smooth part \cite{bb,rs}, whereas the oscillatory shell
structure may be described by semiclassical methods \cite{bb}. In practice,
these two contributions may be well separated in a grand canonical scheme, where the 
energy is considered as a function of the chemical potential $\mu$: the TF 
contribution describes the smooth dependence on
chemical potential, whereas the oscillatory component has {\sl zero average}
(with respect to $\mu$) and describes deviations with respect to the mean
behavior.

In isolated systems with a well defined number of particles $N$, things are
different. When the dependence of the ground--state energy $E (N)$ is considered
as a function of $N$, one can show \cite{clmrsv} that its fluctuating part, 
$\et (N)$, in contrast to usual expectations, has a non--zero average 
(as a function of $N$). It follows that the fluctuating part contributes to the smooth 
part $\eb (N)$ of the energy, and the frontier between smooth and fluctuating components is
blurred. This is a generic effect, although its importance depends on the symmetries 
(and intrinsic dynamics) of the mean field potential. A description of this effect 
was recently developed \cite{clmrsv} for a confining potential which keeps its shape fixed 
(up to a possible scaling factor) when the number of particles varies (cf~Eq.~(\ref{eq105})~below). 

Although the latter situation may be relevant in many experimental set--ups, like 
cold dilute Fermi gases in magnetic atom traps \cite{hara}, where the external HO potential
dominates over the mean--field interaction energy \cite{gwo}, another relevant case is
that of self--bound systems, like the atomic nucleus. In these systems, an effect that
was not included in the previous description appears: the shape of the average
self--bound confining potential depends on the number of particles. As is well 
known, at a given $N$ the shape is determined by minimizing the energy of the system. 
The minimization of the smooth part of the energy leads generically to an 
isotropic shape. We denote this contribution $\eb_{sph} (N)$.
For that shape, and for particular values of $N$ (magic numbers)
the contribution of the fluctuating part $\et_{sph} (N)$ is large and negative,
thus reducing the total energy with respect to $\eb_{sph} (N)$. Away from magic numbers,
the amplitude of $\et_{sph} (N)$ rapidly decreases, and may eventually become positive. 
In order to avoid this behavior, the system deforms, trying to keep the value of the oscillating 
part of the self--bound energy $\etsb (N)$ negative and as large as possible.
Though in most realistic situations $\etsb (N) \ll \eb (N)$, the behavior of $\etsb (N)$ has
a strong influence on the shape of the system. Schematically (see Fig.\ref{fig6}), the behavior 
of $\etsb (N)$ is therefore a fluctuating negative function of $N$, with
larger amplitudes around the magic numbers. 

It follows that in self--bound systems, as for Fermi gases confined by a fixed external 
potential, the average part of the energy fluctuations is again generically
different from zero. However, the properties of the average are very different in these two cases.
In particular, the average is positive for a fixed shape, while it is negative
in self--bound systems.
Such a bias toward negative energies of the fluctuating part in self--bound systems 
is clearly observed in realistic calculations \cite{clmrsv}. In the bottom part of 
Fig.~\ref{fig7} we plot the nuclear ground--state shell correction energy computed in
Ref.\cite{mnms} using a macroscopic--microscopic model, whose results are in good
agreement with experimental data. We do observe a
tendency toward negative values, with a non-zero slope for the average part
of the fluctuations. One of our purposes here is to provide a 
quantitative description of this effect.

We will consider the case of non--interacting fermions in 3D whose
self--bound confining potential is assumed to have a quadratic harmonic shape.
The reason for such a choice is purely technical, since it allows, to some
extent, for an explicit analytic description. In spite of its simplicity, we
will show that it provides a correct description of what is observed
in more realistic calculations.

Harmonic potentials were intensively investigated
in the past, starting from the Nilsson model of the nuclear deformation
\cite{nilsson}. This is an integrable (separable) model. It is known that the
statistical properties of its single--particle spectrum do not coincide with
the generic (Poisson) behavior expected in integrable systems \cite{pbg}. The 
statistics, explored mainly in D=2, are in fact very sensitive to the number 
theoretic properties of the frequency ratios \cite{pbg,bj}. As we will show,
these number theoretic properties also strongly influence the behavior of
the many body system.

Our analysis of the minimization of the energy of the Fermi gas will be based 
on a semiclassical theory. Though this theory is exact for the single--particle 
density of states, the different cases (of irrational or rational frequency ratios) 
should, however, be considered carefully \cite{bj}. 
We shall see that the amplitude and phase of shell effects 
are directly controlled by the number--theoretic properties of the frequencies. 
The output of the minimization problem thus depends on a delicate 
interplay between these number--theoretic properties and the number of
particles in the gas.

The manuscript is organized as follows:
in section II we study the main properties of
$\et (N)$ for an harmonic potential of given frequency ratios. We will illustrate
the large variety of different behaviors that could emerge (including shell and
supershell structures). In section III, we deal with the case of
self--bound systems, and apply our results to a schematic model of the
atomic nucleus (that we compare to realistic calculations).

\section{I. General Setting}

For a given potential 
the single--particle level density is decomposed into a smooth 
part $\rhob$ coming from the TF theory plus an oscillatory contribution
$\rhot$:
\begin{equation}\label{eq101} 
\rho (\epsilon) = \rhob (\epsilon) + \rhot (\epsilon) \ .
\end{equation}
Here, $\epsilon$ is the single--particle energy. Below, we will show in more 
detail how to describe these two components. The spin
degeneracy is included in the level density. Equation (\ref{eq101}) induces a 
corresponding decomposition of the integrated level density (that counts the number of
single--particle states up to an energy $\mu$):
\begin{equation}\label{eq102}
\nc (\mu) = \ncb (\mu) + \nct (\mu) 
=\int^{\mu}_0 \rhob(\epsilon) \, d\epsilon \ 
+ \int^{\mu}_0 \rhot(\epsilon) \, d\epsilon \ .
\end{equation}
In order to study a system with a finite number of
particles, we use canonical expressions for
thermodynamic quantities. For a system of $N$ non--interacting fermions, 
we define its ground state energy $E(N)$, the shell correction energy $\et (N)$
and the smooth TF component $\eb (N)$ as \cite{bb,rs}:
\begin{equation} \label{eq103}
\et(N) = E(N)-\eb(N)
= \int_0^{\mu_{_{\scriptscriptstyle N}}} \epsilon \rho(\epsilon) d\epsilon
- \int_0^{\mub_{_{\scriptscriptstyle N}}} \epsilon \rhob(\epsilon) d\epsilon .
\end{equation}
The chemical potential $\mu_{\scriptscriptstyle N}$ and its smooth part
$\mub_{\scriptscriptstyle N}$ fix the number of particles.
They are defined by inversion of the exact and average integrated level densities, 
$\nc (\mu_{\scriptscriptstyle N}) = N$ and 
\begin{equation} \label{ncb} 
\ncb (\mub_{\scriptscriptstyle N}) = N \ ,
\end{equation}
respectively. Unfortunately, Eq.(\ref{eq103}) is difficult to exploit 
analytically because the discretization of
$\mu_{\scriptscriptstyle N}$ is difficult to impose. From Eq.(\ref{eq103}) it can
be shown that, neglecting terms of second order in the parameter $\mu - \mub$, $\et$ 
may be approximated by \cite{mbc,lm}:
\begin{equation} \label{eq104}
\et(N) \approx  - \int_0^{\mub_{_{\scriptscriptstyle N}}} \nct(\epsilon) d\epsilon \ .
\end{equation}
This, together with the definition of $\eb (N)$, are the basic equations upon which 
our analysis of ground state energies of Fermi gases will be based on.

If, in Eq.~(\ref{eq104}), $\mub_{\scriptscriptstyle N}$ is considered as a continuous 
variable, it can easily be shown that it gives a wrong result for the
average of the fluctuating function $\et (N)$.
Indeed in a system with a fixed number of particles, the chemical potential
takes discrete values as $N$ varies.
The fluctuating part of the ground state energy is sampled at 
particular values of the chemical potential, implying a modification of 
its average value. 
Recently, an explicit description for this effect was given.
It was found that the contribution of the fluctuating part to the average of 
the energy is given by \cite{clmrsv} 
\begin{equation} \label{eq105}
\etmean \approx \langle \nct^2 (\mub_{\scriptscriptstyle N})
\rangle_{\mub_{_{\scriptscriptstyle N}}}/\rhob(\mub_{\scriptscriptstyle N}) \ ,
\end{equation}
where we use brackets to denote an average over an appropriate chemical potential window
and to distinguish this contribution with respect to the TF smooth term.

\section{II. External potentials}

\subsection{A. Triaxial case with irrational frequencies}

We consider a particle in a 3D harmonic potential. The Hamiltonian is given
by:
\begin{equation} \label{eq201}
H = \frac{1}{2} (p_x^2+p_y^2+p_z^2) +\frac{1}{2} (Q_1^2 x^2+Q_2^2 y^2+Q_3^2 z^
2) \ ,
\end{equation}
where all quantities are dimensionless (units $\hbar=m=1$).
$Q_1$, $Q_2$ and $Q_3$ are the frequencies of the harmonic oscillator (HO).
In this subsection we considered them as incommensurable real numbers with 
pairwise irrational ratios. 
Using Eq.(\ref{eq102}) and the expression of the TF level
density $\rhob$ given in Table I,
we calculate from Eq.(\ref{ncb}) the smooth chemical potential
to leading order in $N$: 
\begin{equation} \label{mub}
\mub_{\scriptscriptstyle N}=(3 Q_1 Q_2 Q_3 N)^{1/3} \ .
\end{equation}
For irrational frequency ratios, the only periodic orbits of the system are the 
one--dimensional oscillations along the three principal axis $x$, $y$ and $z$.
These orbits are isolated, and are the backbone for the description of the fluctuating
part of the different quantities.
In particular, the single particle level density $\rhot (\epsilon)$ was computed
in \cite{bj} (using semiclassical methods based on the trace formula, which 
coincide with exact results obtained from the inverse Laplace transform
of the exact partition function). From Eqs.(\ref{eq102}) and (\ref{eq104}), 
and the previous expression (\ref{mub}) of $\mub_{\scriptscriptstyle N}$, 
we compute analytically the expressions for $\nct(\mu)$ and $\et(N)$. These, together with that of
$\rhot$ (see Ref.\cite{bj}), are reported in Table II (first column, case A). 

\begin{figure}[h]
\resizebox{1.0 \linewidth}{!}{\input{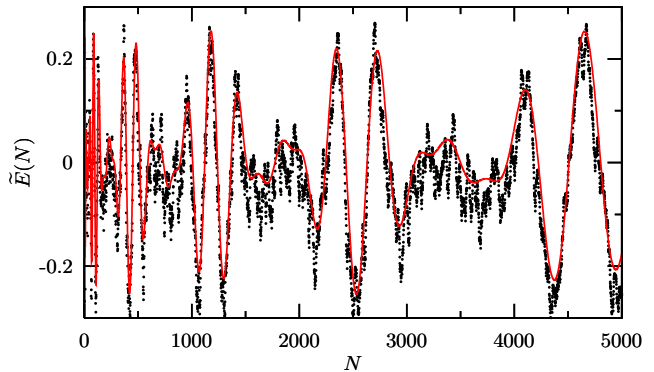}}
\caption{\label{fig1} 
(Color online) Ground state shell correction energy as a
function of $N$ for $Q_1=\sqrt2$, $Q_2=\sqrt3$ and $Q_3=\sqrt5$.
Quantum computation in dots, theoretical prediction (see Table II)
using the first term in each sum in full line.}
\end{figure}

From periodic orbit theory (POT), 
it is known that the main features of the shell correction energy 
can be recovered by taking into account only the first (shortest) orbits \cite{lm}, 
which correspond to
the first terms in each of the three sums (lowest $k$ values) of $\et (N)$. 
As the number of particles $N$ is varied, the interferences between these terms 
produce an oscillatory pattern. For potentials in which the
single--particle spectrum has a simple structure,
like the isotropic HO discussed in the next subsection, the oscillatory pattern 
of the energy as a function of $N$ will be quite regular. One speaks in that case of shell effects
and magic numbers (which correspond to the minima of the regular oscillation). In the present case of
irrational frequency ratios, the structure of the single--particle spectrum is highly
non--trivial, and leads to a complicated pattern of oscillations of the energy as a function
of $N$. This is illustrated for a particular case in Fig.~\ref{fig1}. The full line is obtained
from the corresponding equation of Table II, using only the first term in each sum ($k=1$).
Although details are missing, note the good overall agreement obtained with just three orbits. 
Note the absence of regularity of the pattern.

As the classical orbits for irrational frequencies do not form families but are instead isolated,
the amplitude of the fluctuations is small (compared to the results of, e.g., the isotropic case
of the next subsection, see Fig.~\ref{fig2}). The purely quantum mechanical counterpart
of this statement is that due to the absence of symmetries in the system, no systematic degeneracies 
occur in the single--particle spectrum, fluctuations are small, and accordingly the energy does not deviate 
significantly from its average part. Moreover, the typical value of 
$\nct$ is also small, and we find a correction to the mean value coming from the oscillatory part,
calculated from Eq.(\ref{eq105}), that vanishes as $N^{-2/3}$ (cf
\cite{clmrsv} for further details of the method).

\subsection{B. Isotropic case}
\begin{figure}[t]
\resizebox{1.0 \linewidth}{!}{\input{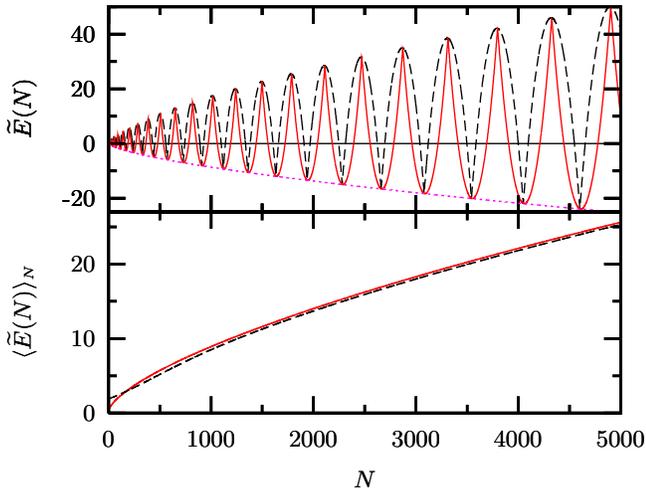}}
\caption{\label{fig2} 
(Color online) Upper panel: Ground state shell correction
energy as a function of $N$ for the isotropic HO. Quantum computation
in dashed line, semiclassical theoretical prediction in full line. The dotted line
correspond to the analytical expression of the minima of $\et$ (see last row of
Table II).\\
Lower panel: Numerical average of the exact fluctuating function of the upper panel
(dashed line), compared to the theoretical prediction,
Eq.(\ref{eq105}) (full line) (from Ref.\cite{clmrsv}).}
\end{figure}

Let's consider now the case $Q_1=Q_2=Q_3=1$. The system possesses the U(3) symmetry. Within POT,
a perturbative approach of the irrational case allows to compute the level density 
\cite{bj}, to obtain the well known result given in the top row of the second column of 
Table II.
To compute $\nct(\mu)$ and $\et(N)$, we have
followed the method explained above (see the results in the second column of Table II). The high
degree of symmetry of the system leads in this case to a single characteristic period for the periodic
orbits, implying a much more regular pattern of
the shell oscillations (compare the upper part of Fig.~\ref{fig2} with Fig.~\ref{fig1}). With respect to
the triaxial irrational case, the amplitude of the fluctuations is much larger 
(this is related to the high degeneracy of the energy levels or, semiclassically,
to the fact that the periodic orbits form families). The approximate frequency of the shell fluctuations is
given by the phase of the cosine function of the first $k=1$ term of the sum in
$\et(N)$. Thus magic numbers, given by the values of $N$ that minimize $\et(N)$, are well 
approximated by 
\begin{equation}\label{eq202}
N_{\scriptscriptstyle MAGIC}=\frac{i^3}{3}, \;\;\;\;\; \mathrm{where} \ i \in \mathbb{N^{*}} \ ,
\end{equation}
which corresponds to integer values of the chemical potential $\mub_{\scriptscriptstyle N}$.

Classically, all the orbits are closed (periodic), and have the same period. This high degeneracy
of the periodic orbits is reflected in the prefactor $N^{2/3}$ in front of
$\et (N)$ (cf Table II), as compared to the case of irrational ratios. Because there is only one 
characteristic frequency associated to the periodic orbits, no beating
effects are observed, and the fluctuations are very regular (forming "shells"). 
In particular, no supershell structure is present, like the characteristic
beating pattern observed in a spherical cavity with hard walls, produced by 
the interference between the triangular and the square orbits, see Refs.\cite{bb,nhm}.

In Fig.~\ref{fig2}, we clearly see that the average of $\et(N)$ is non zero and positive, 
and increases as $N$ increases. Using in Eq.(\ref{eq105}) the corresponding expression 
for $\nct(\mub_{\scriptscriptstyle N})$, we get an
analytical expression for $\etmean$ (cf fourth row of Table II, and \cite{clmrsv} for 
further details). 
Theory and numerics are compared in the bottom part of that figure.

\subsection{C. Triaxial case with rational frequencies}
Now $Q_1$, $Q_2$ and $Q_3$ are positive integers and have pairwise irreducible
ratios.
\begin{figure}[ht]
\resizebox{1.2 \linewidth}{!}{\input{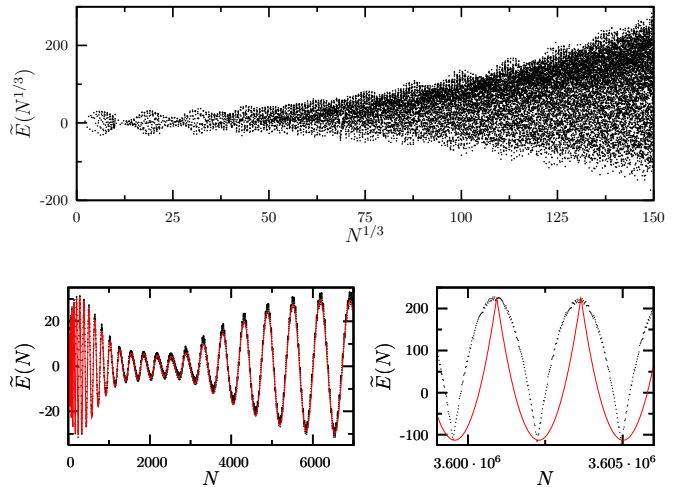}}
\caption{\label{fig3} 
(Color online) Upper panel: 
Ground state shell correction energy as the function of
$N^{1/3}$ for $Q_1=17$, $Q_2=18$ and $Q_3=19$. In
dots, numerical computation. We observe 
triaxial to isotropic transition for $\rct \backsim 1$, which
corresponds to $N_c~=~131060$ (or $N_c^{1/3}~\backsim~51$).\\
Lower left panel:
Magnification of the upper panel for low values of $N$ ($\rct ~>~
7$). In dots, numerical computation. The full line corresponds to
the analytical expression of $\etA$ with the first term in each sum.
The triaxial component of $\et$ dominates, and shows super
shell effects due to interferences of the shortest orbits.\\
Lower right panel: 
Magnification of the upper panel for $N^{1/3}~\backsim~150$ 
($\rct~ \backsim~ 0.11$).
Dots, numerical computation. The full line corresponds to
the analytical expression of $\etS$.
The isotropic component of $\et$ dominates. The behavior is similar
to the isotropic case of Fig.~\ref{fig2}.
}
\end{figure}

Since frequencies are integer, it produces families of classical periodic
orbits (Lissajous figures). A perturbative treatment, similar to the isotropic case,
can be done to obtain an explicit expression for the level density \cite{bj}. 
For convenience, the oscillating part of the level density is decomposed 
into two different parts (see Table II): $\rhotI$ (resp. $\rhotT$) which 
is connected to the level density of the isotropic
(resp. triaxial with irrational frequency ratios) HO. Below, these two components
are referred to as "isotropic" and "triaxial", respectively. We follow the same
method as before in order to get expressions for $\nct(\mu)$ and $\et(N)$ 
(cf third column of Table II).

To understand in this case the $N$--dependence of the fluctuating part of the energy, 
it is useful to calculate the ratio 
\begin{equation}\label{ratio}
\rct =
\frac{\Lambda_{\scriptscriptstyle T}} {\Lambda_{\scriptscriptstyle I}}
\end{equation}
of the typical amplitude of the triaxial component with respect to the isotropic one.
These two amplitudes are approximated here by computing the amplitude of the
first term $k=1$ of the corresponding sum (cf Table II)
{\small\begin{eqnarray}
\Lambda_{\scriptscriptstyle I}&=&\frac{(3 N)^{2/3}}{2 \pi^2 (Q_1 Q_2
Q_3)^{1/3}}\label{eq203} \ , \\
&&\notag\\
 \Lambda_{\scriptscriptstyle T}&=&\max \left( \begin{array}{c}
Q_1/(4\pi^2|\sin(\pi Q_2/Q_1)\sin(\pi Q_3/Q_1)|)\\
Q_2/(4\pi^2|\sin(\pi Q_1/Q_2)\sin(\pi Q_3/Q_2)|)\\
Q_3/(4\pi^2|\sin(\pi Q_1/Q_3)\sin(\pi Q_2/Q_3)|)
\end{array} \right). \label{eq204}
\end{eqnarray} }
A finer estimate of $\Lambda_{\scriptscriptstyle T}$ requires an 
analysis of the number theoretic properties
of the $k$-th dependence of the denominators in $\et_{\scriptscriptstyle T}$, 
that we shall not do here. Since, for given frequencies, 
$\Lambda_{\scriptscriptstyle T}$ is a constant independent of $N$, 
$\rct \propto  N^{-2/3}$. Thus, for any rational set of frequencies, there exists
a critical number of particles $N_c$ above which the behavior
of $\et(N)$ is dominated by the isotropic component. In this high--$N$ regime the 
behavior of $\et (N)$ qualitatively
coincides, up to a rescaling of the overall amplitude by $(Q_1 Q_2
Q_3)^{-1/3}$ and of a rescaling of the phase by $(Q_1 Q_2 Q_3)^{1/3}$,
with the behavior of the isotropic HO
described in the subsection II.B. This corresponds to a regular (i.e. single--frequency) 
pattern of large amplitude 
oscillations (or shells), with clearly defined magic numbers,
as a function of $N$ (compare Fig.~\ref{fig2} 
and the lower right panel of Fig.~\ref{fig3}). The value of the magic numbers
may be estimated similarly to the previous subsection. This behavior illustrates 
the occurrence of an increasing number of accidental
degeneracies in the single--particle spectrum as $N$ increases 
(as already pointed out in Ref.\cite{pbg} in the 2D case).
This qualitative behavior is also valid for the axial symmetric case, where two
integer frequencies are equal.

In contrast, in the limit $\rct \gg 1$ ($N \ll N_c$), $\et (N)$ is
controlled by the triaxial term of the shell energy, $\etA$.
As in the irrational case, the fluctuating part of the energy is
now characterized by a roughly constant (and small, compared to the isotropic case)
amplitude, with the gross features of the oscillations well approximated by the 
superposition of a small number of terms having different frequencies. In particular,
for some triplet of frequencies, in the regime $N < N_c$ supershell-like structures may be observed,
as illustrated for instance in Fig.~\ref{fig3}.

In summary, in the triaxial rational case generically a transition at $\rct \sim 1$ 
($N \sim N_c$) from a triaxial regime ($\rct \gg 1, N \ll N_c$) toward an isotropic behavior 
($\rct \ll 1, N\gg N_c$) 
is expected as $N$ increases. This behavior is illustrated for a particular
set of frequencies in Fig.~\ref{fig3}, where the transition $\rct \sim 1$, estimated
from Eqs.(\ref{ratio})--(\ref{eq204}), occurs 
at $N_c \sim 131060$. The lower left and right panels of Fig.~\ref{fig3}
focus on the two extreme limits. 

Though we have not studied all the possible cases (axial symmetry, rational-irrational, etc),
the above examples illustrate the variety of behaviors of energy fluctuations one can find as the
frequency ratios are varied.


\section{III. self--bound systems}
Up to now we have considered non--interacting fermions in HO
potentials that keep their shape (frequencies) fixed as the number of fermions is varied.
Such a scheme is a good approximation for dilute Fermi gases in optical traps
\cite{gwo}, or in quantum dots, where the external potential dominates over the mean--field 
interaction energy \cite{gwo}. But self--bound systems, like metallic clusters or
atomic nuclei, behave differently. In those systems the shape of the self--consistent
mean field potential, determined from the minimization of the ground 
state energy with respect to deformations, may strongly depend on the number of particles
$N$.

If the HO potential considered here is viewed as the self--consistent mean field potential, 
then at each $N$ the energy is minimized with respect to the three frequencies. 
Conservation of the volume implies that the product of the three
frequencies remains constant \cite{bm}, $Q_1 Q_2 Q_3 =cte$. 
The constant can be in fact a function of $N$, to mimic for instance the approximately constant 
nuclear density. But this acts as a multiplicative factor
which can be absorbed into the frequencies. Hence, without loss of
generality, we normalize this constant to one.

In the following we consider only axial
symmetric deformations ($Q_1 = Q_2$). This simplification, though not exact, is known 
to provide a good approximation in most nuclear cases. We use the axially symmetric expression 
of the oscillating part of the energy computed analogously to the previous sections, 
with $Q_1$ and $Q_3$ two incommensurable real numbers with pairwise irrational ratio, and imposing the
volume conservation. There is thus only one free parameter ($Q_1$), and we obtain
\begin{eqnarray}
&\etsb (N)&= \frac{(3 N)^{1/3}}{2 \pi^2} \sum_{k = 1}^{\infty}\frac{\sin(2 \pi k (3
N)^{1/3}/Q_1)} { k^2  \sin(\pi k / Q_1^{3})}\notag\\
&+& \frac{1}{4 \pi^2 Q_1^2}  \sum_{k =1}^{\infty} \frac{\cos(\pi k
/Q_1^3) \cos((2 \pi k (3 N)^{1/3}/Q_1))}{k^2\sin^2(\pi k
/Q_1^3)}\notag\\
&+& \frac{1}{4 \pi^2 Q_1^2} \sum_{k = 1}^{\infty}\frac{(-1)^{k+1}
\cos((2 \pi k (3 N)^{1/3} Q_1^2))}{k^2 \sin^2(\pi k Q_1^3)} \ . \label{eq300}
\end{eqnarray} 
Using similar arguments as in previous sections, we keep only the first term in each
sum of Eq.(\ref{eq300}). Besides, as we must minimize the energy with respect to
$Q_1$, the latter is a function of $N$. The term which comes from the
first sum has the largest amplitude (this will be confirmed {\sl a
posteriori}, when the $N$--dependence of $Q_1$ will be know) \cite{irr_vs_r}.
So we can remove all other contributions from Eq.(\ref{eq300}), keeping
the simple approximation
\begin{equation}\label{eq301}
\etsb(N) = (3 N)^{1/3} \frac{\sin(2 \pi \ (3 N)^{1/3}/Q_1)} 
{2 \pi^2  \sin(\pi / Q_1^3)} \ .
\end{equation} 

\begin{figure}[h]
\resizebox{.90 \linewidth}{!}{\input{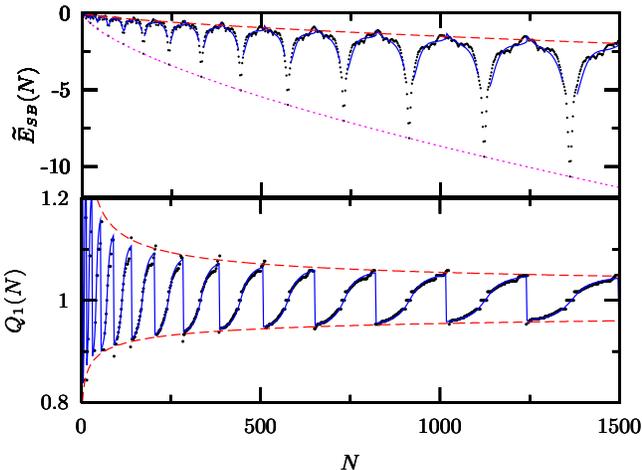}}
\caption{\label{fig6}(Color
         online) Upper panel: Oscillating part of the ground state energy as a
function of $N$ computed numerically by the minimization of the energy of the free
triaxial HO (dots). In full line we plot Eq.~(\ref{eq301}) using the
phenomenological expression (\ref{eq305}) for $\varepsilon$. In
dashed line, analytic result for $\etsb$ valid at mid--shells, Eq.(\ref{eq304}). 
In dotted line, analytic result for $\etsb$
for the magic numbers (Table II, case B, last row). \\
Lower panel: $Q_1 = (1+\varepsilon)^{-1/3}$ as a function of $N$
computed numerically by the minimization of the energy of the free triaxial HO (dots).
In full line we plot $Q_1$ using the phenomenological expression (\ref{eq305}) for
$\varepsilon$.
In dashed line, the mid--shell approximation, Eq.(\ref{eq303}).}
\end{figure}

For the minimization, we need also to evaluate $\eb (N)$. The approximated
(leading--order) TF chemical
potential Eq.(\ref{mub}) used until now is not sufficiently accurate because
it leads to a wrong TF ground state: the minimum of $\eb$ doesn't give an isotropic shape.
To obtain a better description of $\eb$ we must improve the calculation of 
$\mub_{\scriptscriptstyle N}$. This implies solving the full cubic equation given by the canonical 
condition $\ncb (\mub_{\scriptscriptstyle N}) = N$. We have done that using Cardano's
formula \cite{as}, and obtained the result presented in the third row of Table~I.
Then, the analytic expression for the energy to be minimized is
\begin{eqnarray}
&&E(N)= \frac{(3 N)^{4/3}}{4}+\frac{(3 N)^{2/3}}{24} \theta^{-2/3}
(2+\theta^2)
\notag\\
&+&\frac{\theta^{-4/3}}{2(12)^2} (2+\theta^2)^2
+\frac{(3 N)^{1/3} \sin(2 \pi (3 N \theta)^{1/3})}{2 \pi^2\sin(\pi
\theta)} \ , \label{eq302}
\end{eqnarray}
where
\begin{equation}
\theta =  Q_3/Q_1 = Q_1^{-3} \ . \label{eq3022}
\end{equation}

We assume now that the frequency ratio has the perturbative form $\theta~=~1~+~\varepsilon
\, \mathrm{, where} \ | \varepsilon | \ll 1 $ (small deformations).
Since we know that close to magic numbers the system is spherical, we are interested
in the behavior in--between shells. We thus concentrate our analysis of shell effects
for values of $N$ close to the middle of the shells. 
From Eq.(\ref{eq302}), we find that the middle of the shells correspond
approximatively to values of $N$ given by $(3N)^{1/3} = (2 i +1)/2 \  \mathrm{, where} \ i \in
\mathbb{N}$.
Hence for these values of $N$, using a Taylor expansion of
$\theta^{1/3}$ and the addition formula of the sine function, we get
for the fluctuating part of the energy:
\begin{equation}\label{eq3023}
\etsb(N) = \frac{(3 N)^{1/3}}{2 \pi^3 \varepsilon} \sin \bigg( \frac{2 \pi (3 N)^{1/3} 
\varepsilon}{ 3}\bigg) \ .
\end{equation} 
The latter equation is the leading order in
$\varepsilon$ of Eq.(\ref{eq302}) (the smooth part is of lower order). 
The minimization of the energy is thus
equivalent to the minimization of the function (\ref{eq3023}). 
We cancel the derivative of Eq.(\ref{eq3023}), to get
\begin{equation}\label{eq3024}
\frac{2 \pi (3 N)^{1/3} \varepsilon}{3} = \tan\bigg(\frac{2 \pi (3 N)^{1/3} \varepsilon}{3}\bigg) \ .
\end{equation} 
Note that $-\varepsilon$ is also a solution of Eq.(\ref{eq3024}) (prolate--oblate symmetry). 
In spite of the fact that Eq.(\ref{eq3024}) has many solutions, we keep of course the one that gives 
the smallest value of the function (\ref{eq3023}).
We have solved numerically this equation for $\varepsilon$, and find that 
the solution is well approximated by
\begin{equation}
\varepsilon(N) =\pm 1.49 N^{-1/3} \label{eq303} \ ,
\end{equation}
which gives, for the self--bound energy fluctuations at mid--shell
\begin{equation}
\etsb(N) = - 0.0152 N^{2/3} \label{eq304} \ .
\end{equation}

We have tested numerically these results (cf Fig.~\ref{fig6}). They give a very
good approximation of $\etsb(N)$ in the mid--shell region,
showing that the axially symmetric deformation is a good approximation. The weight
of the fluctuating term at mid--shell is thus proportional to $N^{2/3}$, with a negative slope. 
This has the same $N$--dependence as the amplitude of the fluctuating part of an isotropic
HO (whereas an axial--symmetric fixed potential yields instead a $N^{1/3}$ dependence),
or to the subleading, surface term of the smooth part. 
This enhancement shows the fundamental importance of the $N$--dependence of the frequencies. 

Beyond this simple approximation, we have also found a phenomenological expression of 
$\varepsilon$ which turns out to be a quite good approximation of the deformation for
arbitrary $N$ (see bottom part of Fig.~\ref{fig6})
{\small\begin{equation}\label{eq305}
\varepsilon(N) = - \frac{ \pi}{2 (3N)^{1/3}} \arctan \bigg(2 \pi ((3N)^{1/3} - [(3N)^{1/3} +1/2]) \bigg ),
\end{equation}}
where $[a]$ denotes here the integer part of $a$.
This equation shows that the deformation is, up to a global damping factor $\propto N^{-1/3}$,
a strictly periodic function of the variable $x=(3 N)^{1/3}$, of period $\Delta x = 1$
(the same is valid for the shell correction energy).
It gives an extremely accurate description of the $N$ dependence
of the frequency, $Q_1 = [1+\varepsilon (N)]^{-1/3}$, see Fig.~\ref{fig6}.
Increasing the particle number $N$ from the value of a magic number (corresponding to $Q_1 = 1$),
the frequency initially increases, $Q_1 > 1$, corresponding to a prolate shape (accordingly, 
$Q_3 < 1$). At mid-shell,
the shape suddenly changes from prolate, $Q_1 > 1$, to oblate, $Q_1 < 1$, keeping its deviation 
from isotropy $|Q_1 - 1|$ constant. As $N$ further increases, $Q_1$ diminishes, 
to arrive finally at $Q_1 =1$ again at the next magic number. The cycle starts again, with
a decreasing overall amplitude.

Replacing Eq.(\ref{eq305}) in Eq.(\ref{eq301}) we obtain a rather good approximation of
$\etsb (N)$, see Fig.~\ref{fig6}. Using the approximation
$\varepsilon \ll 1$, $\etsb (N)$ can be written
\begin{equation}\label{eq305b}
\etsb(N) = \frac{(3 N)^{2/3} \sin\bigg( 2 \pi \ (3 N)^{1/3} (1 + \varepsilon(N))^{1/3}\bigg)} 
{\pi^4 \arctan \bigg(2 \pi ((3N)^{1/3} - [(3N)^{1/3} +1/2]) \bigg )},
\end{equation} 
where $\varepsilon (N)$ is given by Eq.(\ref{eq305}). The results obtained with this equation
are very similar to those of Eq.(\ref{eq301}), shown in Fig.~\ref{fig6}. This equation clearly
shows the scaling $N^{2/3}$ of $\etsb (N)$. Both the numerator and the denominator in 
Eq.(\ref{eq305b}) are functions of $N$ that oscillate around $0$. The completely 
different behavior of $\etsb (N)$ 
(a negative decreasing function, see Fig.~\ref{fig6}) comes from a delicate balance between
both oscillatory functions. Close to magic numbers, the amplitude of Eq.(\ref{eq305b}),
not shown in the figure, is smaller than the numerical results; further corrections need to be 
included to improve in this region.

We can check that the amplitude $\Lambda_1$ of Eq.(\ref{eq305b})
(first term of the first sum in Eq.(\ref{eq300})) is dominant with respect
to the amplitudes $\Lambda_2$ and $\Lambda_3$ of the first terms of the two remaining 
sums in Eq.(\ref{eq300}), respectively. 
Using $Q_1^{-3} = \theta = 1 + \varepsilon$, the value of $\varepsilon$ 
given by Eq.(\ref{eq303}), and keeping only
the first term of a Taylor expansion in $\varepsilon$ of the sine function
in the denominators, one can easily check that   
$\Lambda_2/\Lambda_1 = \Lambda_3/\Lambda_1 \approx 0.07$. This justifies
the approximation used.

Because the study of $E(N)$ for self--bound system is a minimization problem, we are not
able to compute an explicit and general expression for the average of
the fluctuating part of the energy, analogous to Eq.(\ref{eq105}) but now
valid for self--bound systems. Nevertheless, it follows from the previous results
that the average scales 
as $N^{2/3}$. A numerical fit of the average of $\etsb$, denoting 
\begin{equation}\label{eq306}
\etsbmean = - a_{\scriptscriptstyle fit} N^{2/3} \ ,
\end{equation}
gives $a_{\scriptscriptstyle fit} = 0.0262$.

As an application of the previous results, we consider a schematic model of 
the atomic nucleus,
with $N$ non--interacting neutrons and $N$ uncharged non--interacting
protons whose mean--field potential is assumed to have an HO shape. 
In order to mimic the saturation
properties of nuclear forces and get dimensional quantities we
multiply $\etsb$ by the factor $82 A^{-1/3}$ MeV, where $A = 2 N$
is the mass number \cite{rs}. The previous factor modifies the
$N$ dependence of the amplitude of $\etsb$, leading to an amplitude
now proportional to $N^{1/3}$, instead of $N^{2/3}$. 
The energy is minimized as previously. 

\begin{figure}[h]
\resizebox{1.40 \linewidth}{!}{\input{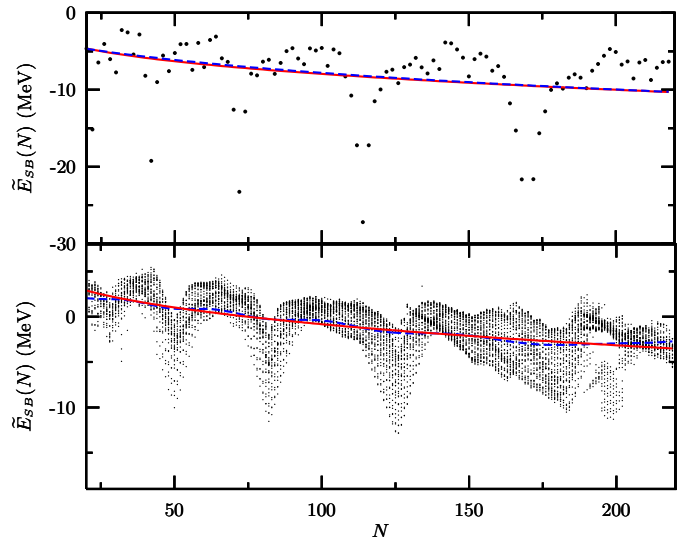}}
\caption{\label{fig7} (Color online) Upper panel: Numerical computation of $\etsb$ as
a function of the neutron number for the HO atomic nucleus model 
(dots). In dashed line, the numerical average of $\etsb$ and in
full line the phenomenological result adapted from Eq.(\ref{eq306}):
$\etsbmean = -~1.71~N^{1/3} \mathrm{MeV}$\\
Lower panel: Ground state shell correction energy
as a function of the neutron number from \cite{mnms} (dots).
In dashed line, the numerical average of $\etsb$ and in
full line the phenomenological fit: $\etsbmean = (8.14-1.93 N^{1/3})$ MeV.}  
\end{figure}

We compare the results of this model to one of the best
theoretical models for the nuclear
binding energy, based on an extension of the liquid drop model
\cite{mnms}, see Fig.~\ref{fig7}.  
The smooth part used in Ref.~\cite{mnms} being different from ours,
it leads to a difference in the offset. In spite of that,
we see that the qualitative properties of the simplified
HO model, including the energy scale, are correct.
In both cases we plot the numerical average of $\etsb$. A negative slope of the fit 
in Ref. \cite{mnms}, consistent with $- a_{sw} N^{1/3}$, is observed, with a proportionality
factor $a_{sw}=1.93 \ \mathrm{MeV}$. It is very similar from what we obtain 
from our simplified model 
($a_{HO} =1.71  \ \mathrm{MeV}$, top part of Fig.~\ref{fig7}).
The shell amplitude as a function of $N$ is
also well reproduced. Nevertheless no supershell structure occurs in our analysis \cite{seville}, 
and magic numbers are badly estimated because we are not 
taking into account, in our simple model, spin-orbit effects \cite{bm}.

\section{IV. Conclusion}
For a Fermi gas treated in the mean--field approximation, we provided here a quantitative 
description of the ground--state shell correction energy. Two different possibilities
were considered, fixed potentials and self--bound systems. We showed
that very different qualitative behaviors may appear in harmonic traps, depending on the
number--theoretic properties of the frequency ratios. Based on a semiclassical theory, we
provided an analytic description of the different behaviors, as well as of the average of the
shell energy. In spite of its simplicity, HO potentials show good qualitative 
agreement with more elaborate models. In self--bound systems, one of our main results is
an accurate analytic description of the deformation and of the shell correction energy
as a function of the particle number.

A few periodic orbits were used to describe $\et (N)$. Although this is in general a good
approximation, in some cases it fails. As discussed in \cite{bj} for $\rhot$, the triaxial 
irrational case with frequencies very close to each other can reproduce, to a good approximation, 
the isotropic case, but only if a large number of terms in the sums is included. Generically, 
the convergence properties of the sums over periodic orbits is a delicate problem, directly 
related to the number--theoretic properties of the frequency ratios. We haven't considered this
problem in its full generality; it clearly deserves a closer inspection, in particular in 
connection with the minimization problem.

We have shown that, in self--bound systems, the properties of the shell correction energy 
scale as $- a N^{2/3}$ (or $- a N^{1/3}$, if saturation properties are taken into account),
where $a$ is some positive constant. Although it compares favorably with realistic nuclear 
models, it should be noted that this behavior is specific of HO potentials. In fact, 
if one considers instead a hard--wall spherical potential, supershell structures will appear, 
and modulations of, e.g., the average of $\etsb (N)$ are expected, as observed in
metallic clusters \cite{knight}. These, however, are
not observed experimentally in nuclear data, due to the limited number of nucleons 
(see, for instance, \cite{seville}). Deviations from an HO confining potential toward
a more steeper one of the Woods--Saxon type are generically expected to be produced by 
interactions. Within a spherical symmetry, these deviations from an harmonic confinement
were shown to lead to supershell effects \cite{yoarb}.
We would like to thank Peter Schuck for usefull discussions.

This work was supported by grants ANR--05--Nano--008--02 and
ANR--NT05--2--42103, by the IFRAF Institute.

\begin{table*}[ht]
\includegraphics[scale=1.]{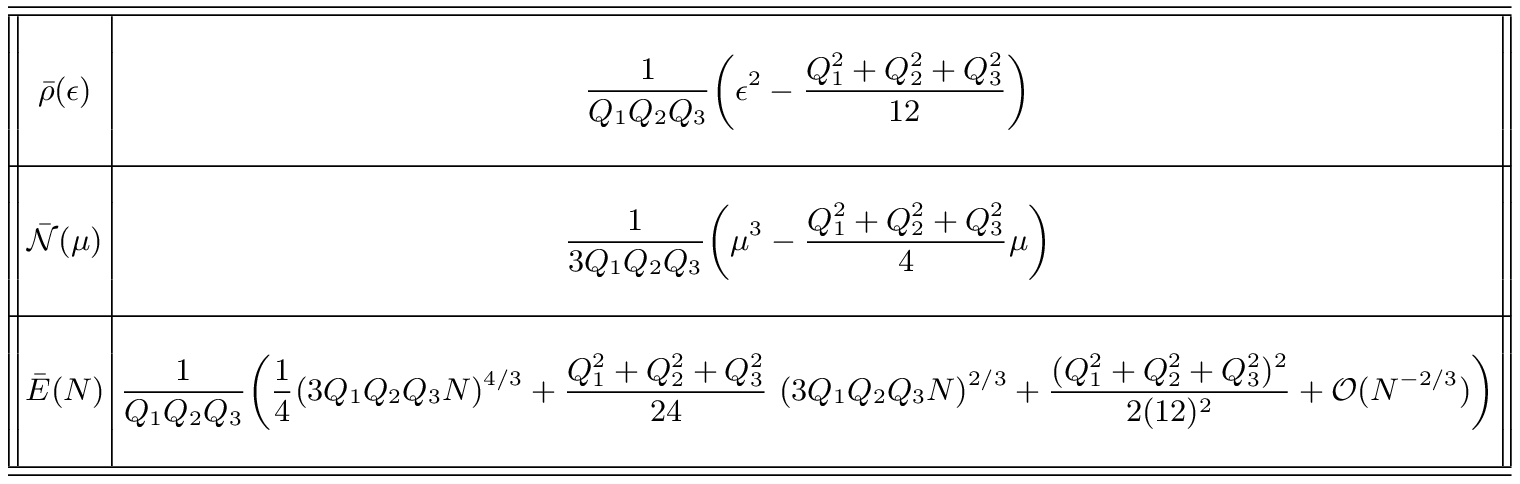}
\caption{\label{table1}
The smooth part of the single--particle level density for a triplet of frequencies 
of the 3D harmonic potential as a function of the energy $\epsilon$ in the first row. 
Corresponding integrated level density up to an energy $\mu$ in the second row. Last 
row, the smooth part of the ground--state energy of the Fermi gas as a function of the 
number of particles $N$.}
\end{table*}
\begin{table*}[]
\begin{narrow}{0in}{0in}
\includegraphics[scale=4.]{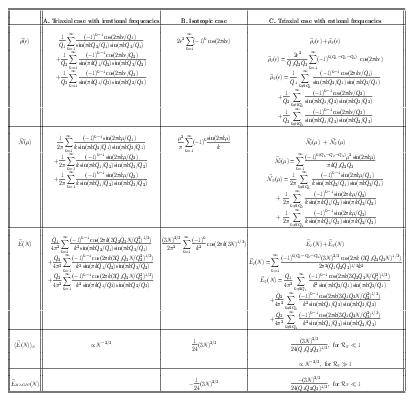}
\end{narrow}
\caption{\label{table2}
The three columns separate the different cases of the 3D harmonic potential studied in 
section II. The oscillating part of the single--particle level density for a triplet of 
frequencies as a function of the energy $\epsilon$ is given in the first row 
(from \cite{bj}). Corresponding integrated level density up to an energy $\mu$ in the 
second row. Third row, the oscillating part of the ground--state energy of the Fermi gas 
as a function of the number of particles $N$. The fourth row gives the analytic expression 
for the average of $\et$. In the last, when available, the analytic curve for the envelope 
of the magic numbers.}
\end{table*}


\begin{thebibliography}{10}
\bibitem{irr_vs_r} We may have chosen a rational frequency ratio between $Q_1$ and
$Q_3$, instead of an irrational one.
However, it can be shown that the isotropic term that appears for a rational ratio, 
discussed in section II.C, is not dominant at the middle of the shells, even in the limit 
$N\rightarrow \infty$, and that the analysis based on rational or irrational frequency
ratios give similar results. This is due to the $N$--dependence of $Q_1$.
\end{thebibliography}
\end{document}